\documentclass[twocolumn,prb,superscriptaddress,showpacs,preprintnumbers]{revtex4}

\usepackage{amsmath,dcolumn,graphicx,hyperref}
\begin{document}

\title{Annealing, lattice disorder and non-Fermi liquid behavior in UCu$_4$Pd}

\author{C. H. Booth}
\email{chbooth@lbl.gov}
\affiliation{Chemical Sciences Division, Lawrence
Berkeley National Laboratory, Berkeley, California 94720, USA}

\author{E.-W. Scheidt} \affiliation{Institut f\"{u}r Physik,
Universit\"{a}t Augsburg, 86159 Augsburg, Germany}

\author{U. Killer} \affiliation{Institut f\"{u}r Physik,
Universit\"{a}t Augsburg, 86159 Augsburg, Germany}

\author{A. Weber} \affiliation{Institut f\"{u}r Physik,
Universit\"{a}t Augsburg, 86159 Augsburg, Germany}

\author{S. Kehrein} \affiliation{Theoretische Physik~III -- Elektronische
Korrelationen und Magnetismus, Institut f\"{u}r Physik,
Universit\"{a}t Augsburg, 86159 Augsburg, Germany}

\date{PRB Rapid Comm. in press, verion 4.3, \today}
\preprint{LBNL-51188}

\begin{abstract}

The magnetic and electronic properties of non-Fermi liquid UCu$_4$Pd 
depend on annealing conditions.  Local structural changes due to this annealing
are reported from U $L_\textrm{III}$- and Pd $K$-edge x-ray absorption
fine-structure measurements.  In particular, annealing decreases the fraction of 
Pd atoms on nominally Cu 16$e$ sites and the U-Cu pair-distance distribution width.
This study provides quantitative information on the amount of disorder in
UCu$_4$Pd and allows an assessment of its possible importance to the observed
non-Fermi liquid behavior.

\end{abstract}



\maketitle


The UCu$_{5-x}$Pd$_x$ system displays many of the canonical behaviors of other
non-Fermi liquid (NFL) systems.\cite{Andraka93}  For instance, at low Pd
concentrations, $x$, it is an antiferromagnet (AF), with the N\'{e}el temperature,
$T_{\textrm N}$, decreasing with increasing $x$.  Near 
$x \approx 1$, $T_{\textrm N}$ is reduced toward zero \cite{Koerner00} and the 
magnetic and electronic properties become that of a NFL: the low-temperature 
linear coefficient of the specific heat and the magnetic susceptibility 
diverge
logarithmically ($\gamma = C/T \propto \chi(T) \propto \log(T)$) or as a
weak power law over a more limited temperature range.\cite{Andrade98}
Unannealed samples also exhibit NFL-behavior
in the resistivity with $\rho(T) - \rho_r \propto T$.  

These observations suggest
a possible quantum critical point (QCP) \cite{Andraka93,Millis93, Continentino96} 
as the cause of the NFL behavior.  Subsequent research 
found that the aforementioned behavior can also
be explained with a Kondo disorder model (KDM), consisting of a broad distribution 
$P(T_\textrm{K})$
of Kondo temperatures, $T_K$, which extends down to very low 
temperatures.\cite{Bernal95,Miranda97b} Inhomogeneous broadening of the copper
nuclear magnetic resonance (NMR) line was taken as evidence for such a 
distribution,\cite{Bernal95} and lattice disorder was identified as its possible 
microscopic
origin:\cite{Booth98c} Pd and Cu atoms were
found to interchange positions within the $C15b$ ($F\bar{4}3m$) lattice, even 
for UCu$_4$Pd, 
which previously was thought to possess a well-ordered structure. 

In spite of the arguments supporting the KDM, there is strong evidence that this
simple model does not contain all the necessary elements to explain the 
experimental data.  Inconsistencies include the anomalously fast muon 
spin-lattice relaxation rates measured 
by muon spin resonance ($\mu$SR) in a longitudinal field,\cite{MacLaughlin00} and the measurements of the
spin-lattice relaxation rate at moderately high fields ($>$51 kOe)\cite{Buttgen00}
from NMR down to 400~mK in UCu$_{3.5}$Pd$_{1.5}$.
In addition, using lattice disorder as 
the microscopic origin for the breadth of $P(T_\textrm{K})$ 
within a tight-binding description for the 
hybridization strength and a constant electronic density of states
has been shown 
to require a continuous, non-thermal distribution in the near-neighbor uranium 
pair distances,\cite{Booth98c} which has not been observed.\cite{Bauer02}  
Other theories that combine critical points with disorder 
phenomena\cite{Castro-Neto98} may eventually explain all
the different data, and recent work considering a
Kondo/quantum spin glass critical point is also promising.\cite{Grempel99}

Despite the evidence against the KDM as a ``minimal model'' describing NFL
behavior in this material, the evidence still indicates that disorder plays an 
important role.  This role is perhaps best exemplified by the series of
experiments\cite{Weber01} on annealed samples of UCu$_4$Pd.
For instance, specific heat 
measurements on annealed UCu$_4$Pd samples show that an AF-like
transition below 200 mK is suppressed toward zero temperature with increasing annealing
time, thereby extending the range of NFL-behavior to lower temperatures.
In addition, the slope of the logarithmic divergence of~$\gamma$ is
slightly decreased by annealing.  
Direct experimental evidence is still required to show whether the main effect of
annealing is to change the degree of Pd/Cu site interchange, and if disorder 
in the U-Cu pair
distance distribution is affected. This latter point is very important to
quantitatively assess the observed changes in, for instance, the specific heat 
data.\cite{Booth98c,Bauer02}

In order to further quantify the role of disorder, we report
x-ray absorption fine-structure (XAFS) experiments 
on samples of UCu$_{4}$Pd at the Pd $K$ and the 
U $L_\textrm{III}$ edges as a function of temperature and annealing time.  We 
find that annealing affects the site interchange (defined as the fraction, $s$, of 
the
total number of Pd atoms to those sitting on 16$e$ sites) and causes a slight ordering 
of the U-Cu pairs, demonstrating that these pairs do exhibit 
some static bond-length disorder, albeit very little.  We consider these results
in terms of various possible models.


The UCu$_4$Pd samples are the same as those reported in Ref. \onlinecite{Weber01} with the addition 
of one sample that was only annealed for one day. We will not discuss the 
splat-cooled sample. 
In particular,
$C/T$ becomes constant below 180 mK in unannealed samples due to 
a possible AF transition,\cite{Koerner00} and the 
Sommerfeld coefficient diverges logarithmically ($\gamma \propto \log (T)$) 
above this temperature.
Annealing in an evacuated quartz tube at 750$^\circ$C lowers the transition 
temperature and the slope of the logarithmic divergence as a function of 
annealing time.  The transition is not observed above 35 mK after 14 days of 
annealing.  Please see Ref. \onlinecite{Weber01} for further details.  

The UCu$_4$Pd samples were prepared for the x-ray absorption measurements
as described in Ref. \onlinecite{Bauer02}.
Samples were loaded into a LHe-flow cryostat, and
absorption measurements were collected in transmission mode on the wiggler
BL 11-2 experimental station at the Stanford Synchrotron Radiation Laboratory (SSRL),
using a LN$_2$-cooled, half-tuned double crystal monochromator.
The vertical collimating slits were set to 0.7 mm for the U-edge data and 0.3 mm 
for the Pd-edge data.
Pd-edge data were collected at $T=$ 20 K and 300 K, although the 300 K data
was only used to verify the low temperature results.  U-edge data were
collected at 20, 100, 200 and 300 K.  Two scans were collected at each temperature.

The XAFS data were reduced and fit in $r$-space as described in Ref. \onlinecite{Bauer02}.
In particular, the XAFS oscillations are
defined as $\chi(k) = \mu(k)/\mu_0(k) - 1$, where $k$ is the photoelectron
wave vector given by $\hbar^2 k^2/2 m_e = E - E_0$, $E$ is the incident photon
energy, $E_0$ is the photoelectron threshold energy as determined by the
edge position, $m_e$ is the bare electron mass,
$\mu(k)$ is the $k$-dependent absorption coefficient, and $\mu_0$ is a
slowly varying background that passes through the XAFS oscillations.  We
determine $\mu_0$ by fitting a 5-knot cubic spline (for the Pd data)
or a 6th-order Chebychev polynomial (for the U data) through the XAFS
above the absorption edge.  The uncertainties in the present data are similar to the higher
quality data in Ref. \onlinecite{Bauer02}.

Fourier transforms of $\chi(k)$ produce peaks that, apart from a calculable
phase shift, correspond to distances between the absorbing atomic species and
its near neighbors.  Fits provide these pair distances, $r_i$, as well as the 
pair-distance
distribution widths, $\sigma_i$, for a given shell of atoms, $i$.
Errors on the fit parameters are estimated from the differences between scans,
a Monte Carlo method \cite{Lawrence01},
and by comparisons to reference materials.\cite{Li95b}  These fits assume a 
site-interchange model\cite{Bauer02} whereby $s$ and the amplitude reduction factor $S_0^2$ (an 
overall XAFS scale factor) determine the amplitudes of all
the scattering shells.  Disorder apart from site interchange is
accounted for only in the measurements of the various $\sigma$'s.
The possible
presence of such disorder is tested by fitting the U-Cu $\sigma$'s to
a correlated-Debye model,\cite{Crozier88} where any observed offset $\sigma_{\textrm{static}}^2$
necessary between the model
and the data is interpreted as static disorder.  No such disorder was
previously observed in the UCu$_{5-x}$Pd$_x$ system within 
0.0004~\AA$^2$.\cite{Bauer02}
In order to extract the U-Cu pair $\sigma$ and $\sigma_{\textrm{static}}$ for
each sample, $s$ needs to be determined
and held fixed for all the temperature data.  Since determining $s$ from the
Pd-edge fits is less dependent on the fitting model (new peaks appear in the
XAFS spectrum), we use the fits to the Pd data to fix $s$ in the U fits.
Values for $s$ are only allowed to vary in the U fits as a consistency check.


\begin{figure}
\includegraphics[width=3.5in, trim=0 20 0 0]{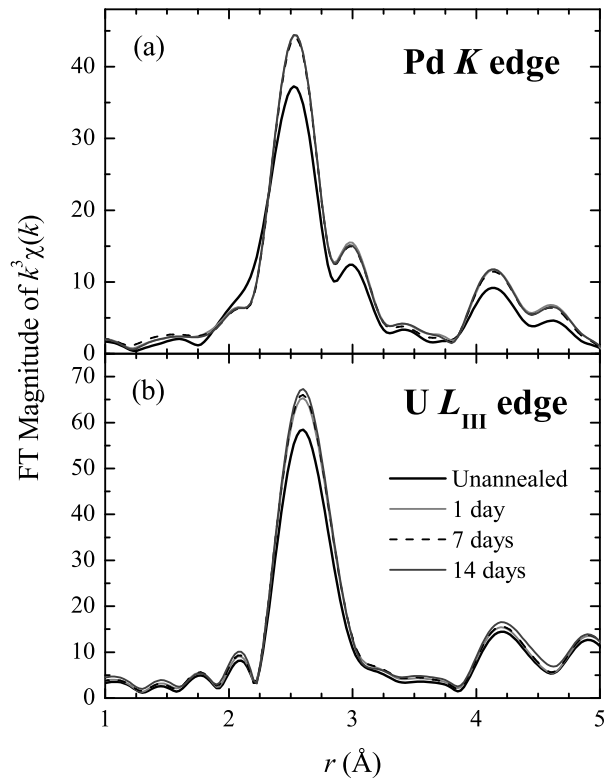}
\caption{Fourier transform amplitudes of $k^3\chi(k)$ XAFS data from the (a) Pd $K$ edge and
(b) the U $L_\textrm{III}$ edge.  Transform ranges are from (a) 2.5-14.8 \AA$^{-1}$ and (b)
2.5-15 \AA$^{-1}$, each Gaussian narrowed by 0.3 \AA$^{-1}$.}
\label{rspace}
\end{figure}

Fourier transform amplitudes of $k^3\chi(k)$ for data collected at 20 K for each
sample are shown in Fig. \ref{rspace}.  The largest effects 
occur in the Pd data.  Here, the overall amplitude increases after the
first day of annealing, \textit{except} in the region near 2.2 \AA~ in the
transform.  This region has the contribution from the Pd$^\prime$-Cu pairs with
$r\approx 2.5$ \AA, where ``Pd$^\prime$'' denotes a Pd on the nominally Cu 16$e$ 
site.  The next peak is mostly due to the nominal Pd-Cu
pairs with $r\approx 2.9$ \AA.  In fact, most of the rest of the spectrum is due to
scattering that occurs in the nominal structure.  Without performing any fits,
these data are explained with the following argument:  After
annealing, some of the site-interchanged Pd$^\prime$ atoms
move onto the $4c$ sites.  This rearrangement removes weight in the 2.2~\AA~ region
of the spectrum due to the relative decrease in the number of Pd$^\prime$-Cu
pairs, while increasing the relative number of pairs involving Pd atoms on 4$c$ 
sites, thereby increasing the amplitude of the rest of the spectrum.  It is
important to note that no measurable changes in the spectra 
occur beyond the first day of annealing.

The U-edge data tell a similar story.
Here, the main difference between the spectra occurs in
the main XAFS peak, primarily due to U-Cu/Pd$^\prime$ scattering at
$\sim$2.9 \AA.  Within this same site-interchange picture, these differences
are due to  U-Pd$^\prime$ pairs converting to U-Cu pairs
on annealing.  The peaks beyond the main peak don't change much because for each
U-Cu/Pd$^\prime$ pair at a given distance, there is a U-Pd pair at a very similar
distance.  A very important difference when comparing to the Pd edge data is that
changes are observed beyond the first day of annealing, which, given the Pd 
edge data, cannot be easily attributed to changes in $s$. 

Fits were performed as described above. In particular, an $S_0^2$ of 0.79(3) was
determined for the Pd-edge fits, and 0.91(3) for the U-edge
fits.  These values are somewhat different than those in Ref. \onlinecite{Bauer02},
probably due to differences in the experimental conditions.
In any case, apart from the derived site interchange values reported below and
pair-width parameters of some further peaks, the
fit results
agree very well with those reported previously, and so we refer the reader to
Tables II and IV in Ref. \onlinecite{Bauer02}.

An important source of systematic error in determining $s$ from the Pd-edge
data is the correlation between the nominal Pd-Cu pairs' distribution
width $\sigma$ and the measured $s$.  This correlation contributes to an
estimated absolute error on $s$ of about 20\%.  However, by
holding this width fixed for all the samples, we obtain a much smaller error
for comparing the differences between the data sets of $\alt 2\%$.
We therefore measure $s=0.27(4)$ for the unannealed sample and
$\Delta s=-0.082(7)$ after and beyond one day of annealing.

Temperature dependent fits to the U-edge data were analyzed using the
correlated-Debye model.  All fits can be described with a correlated-Debye
temperature of $\Theta_\textrm{cD} = 314(4)$ K and
$\sigma_\textrm{static}^2 = 0.0002(3)$ \AA$^2$, indicating little if any
static disorder.  Although this model describes the data well, a systematic 
decrease in the U-Cu pair width with annealing time is observed which can
only be due to changes in a finite $\sigma_\textrm{static}^2$.  Using the data 
collected at 20 K, and defining
$\Delta \sigma^2_X = \sigma^2_\textrm{U-Cu}(\textrm{X})
 - \sigma^2_\textrm{U-Cu}(\textrm{14-day})$,
we measure $\Delta \sigma^2_\textrm{7-day} = 0.00014(5)$ \AA$^2$,
$\Delta \sigma^2_\textrm{1-day} = 0.00029(3)$ \AA$^2$, and
$\Delta \sigma^2_\textrm{0-day} = 0.00044(4)$ \AA$^2$.  These results are, 
therefore,  lower limits on the static disorder in these samples.  Note that a 
changing $\Theta_\textrm{cD}$
cannot explain this ordering, since $\sigma^2 \propto 1/\Theta_\textrm{cD}$ and the
4 K estimated error on $\Theta_\textrm{cD}$ implies a change of only
about 0.00003 \AA$^2$.


\begin{figure}
\includegraphics[width=3.5in, trim=0 25 0 00]{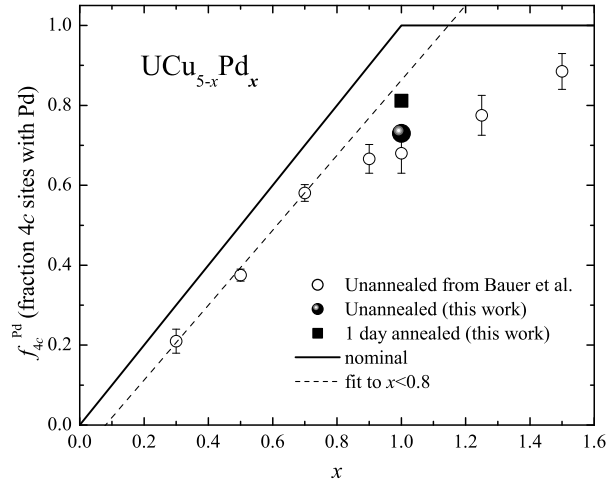}
\caption{The fraction of 4$c$ sites occupied by Pd as a function of $x$ in
UCu$_{5-x}$Pd$_x$.  Data from Bauer \textit{et al}.\cite{Bauer02} is shown for
comparison.  Data for anneals longer than one day are indistinguishable on this 
scale.
}
\label{extrapolation}
\end{figure}

Two important results of the above analysis are that annealing
the UCu$_4$Pd samples reduces the site interchange by about 30\%, and
that even in the sample that was annealed for 14 days, a significant amount of site
interchange still occurs ($s\approx 0.19$).  This latter result
can be inferred from the behavior of $s$ with changing $x$ in 
 UCu$_{5-x}$Pd$_x$\cite{Bauer02}.  Here we follow 
Weber \textit{et al}.\cite{Weber01}, who noted that as $x$ 
increases at low $x$, $a_0$
changes linearly until $x \approx 0.85$, at which time the slope increases.
The difference between the extrapolated line from low $x$ and the
measured $a$ at $x=1$ is about 0.006~\AA.
After annealing the sample, $a_0$ essentially agrees with the extrapolation.  
Bauer \textit{et al}.\cite{Bauer02} noted that the fraction of 4$c$ sites 
occupied by Pd atoms, $f^\textrm{Pd}_{4c}=x(1-s)$, shows very similar behavior as 
the lattice parameter.  To further these comparisons, we show a
similar extrapolation (Fig. \ref{extrapolation}) using the data from 
Bauer \textit{et al}.,\cite{Bauer02} together with the data
collected for the present study.  As in the case of the lattice parameter,
the $f^\textrm{Pd}_{4c}$ of the annealed sample agrees well with the extrapolation for $x < 0.8$.

\begin{figure}
\includegraphics[width=3.5in, trim= 0 25 0 5]{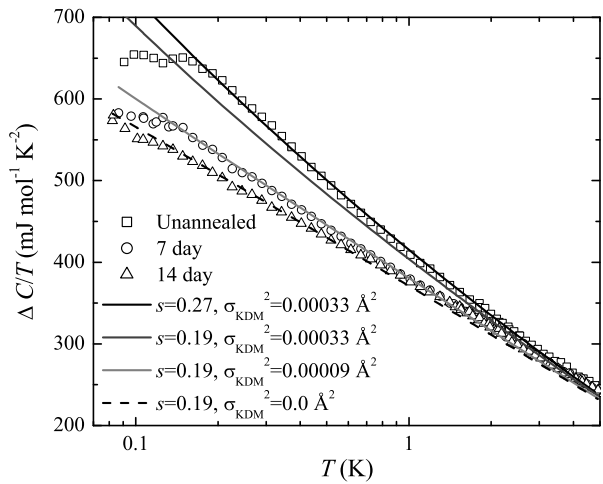}
\caption{The electronic part of the specific heat $\Delta C/T$ as a function of 
temperature from
Weber \textit{et al}.\cite{Weber01}
Fits are with a Kondo disorder model within a tight-binding prescription, as 
described in the
text.}
\label{CoT}
\end{figure}

We will now try to interpret the above results within the
framework of a modified KDM and then within a generic QCP scenario.
Let us first look at the KDM. The specific heat measurements (Fig.~\ref{CoT})
show that the slope of the logarithmic divergence decreases with
annealing time.\cite{Weber01} Qualitatively,
this reduction is expected from a KDM, since the reduced 
lattice disorder should sharpen $P(T_\textrm{K})$, producing 
fewer low-$T_\textrm{K}$ moments and therefore a slower divergence.  This effect
can be calculated using the same tight-binding approach as in 
Refs. \onlinecite{Booth98c,Bauer02}.  However, we reiterate that this model only 
fits the susceptibility and specific heat data if one allows a wide, static 
distribution of U-Cu pair distances, which is not observed in these 
materials.\cite{Bauer02}  Therefore, we are forced to assume that a large 
contribution to the width of $P(T_\textrm{K})$
is due to some other unknown source.  To 
describe this situation, consider a width $W_V$ to the $f$/conduction electron
hybridization energy distribution $P(V_{fc})$ (which generates $P(T_\textrm{K})$) 
as arising from two uncorrelated contributions:
\begin{equation*}
W_V^2=W_0^2+W_\textrm{KDM}^2,
\end{equation*}
where $W_0$ has an unidentified origin and $W_\textrm{KDM}$ is given by a
tight-binding model with the U-Cu static bond length distribution 
width $\sigma_\textrm{KDM}^2$ and
\textit{no adjustable parameters} (see Ref. \onlinecite{Bauer02} for details).  
Using this distribution in a KDM, one can fit the specific heat data
allowing the Fermi energy, $E_F$, the
ratio of the density of states at the Fermi level to the bare $f$-level energy, $\rho/\epsilon_f$,
and $W_V$ to vary while holding $s$ at the measured value,
as shown in Fig. \ref{CoT}.  Fits to the data from each of the samples allow
a single set of most of these variables to be used:
$E_F=1.11$~eV, $\rho/\epsilon_f=0.132$~eV$^{-2}$, and $W_0=0.1448$ eV,  where
$W_0$ is obtained by defining 
$\sigma_\textrm{KDM}^2$(14--day)=0.  These fits then give 
$\sigma_\textrm{KDM}^2$(7--day)=0.00009(3) \AA$^2$ and 
$\sigma_\textrm{KDM}^2$(0--day)=0.00033(3) \AA$^2$, in reasonable agreement with 
the XAFS results.  
We therefore conclude that the KDM requires a significant
unidentified disorder component that is not understood purely in terms of the 
crystal structure, but that the model can be used to describe the \textit{changes} in 
the specific heat very well in terms of the measured structural changes.  
A possible origin of $W_0$ is fluctuations in the local density of 
states,\cite{Miranda01} which are not included in the present model.  
In any case, please
note that we do not
regard $W_0$ as necessarily describing a real width in $P(V_{fc})$, but rather that
this model describes a real contribution ($W_\textrm{KDM}$) to the distribution 
width in addition to some other mechanism.  The other mechanism may only be 
phenomenologically described in terms of a finite $W_0$.  

Another theoretical framework for understanding the NFL-behavior is a
quantum critical scenario. Here one interprets the specific heat data in Fig.~3
as a suppression of an AF transition $T_{\rm N}\rightarrow 0$
due to annealing indicated by the shift in the leveling off of the curves.
The logarithmic behavior in $C/T$ is then a consequence of quantum critical
fluctuations. 
However, this interpretation suffers from the current lack
of an established theoretical
understanding how AF~quantum critical fluctuations can lead to logarithmic
behavior in $C/T$ in a three-dimensional compound, though various
theories have been put forward.\cite{Stockert98,Si01} Finally, we would like
to note in passing that the total entropy quenched by the logarithmic
slope in UCu$_{4}$Pd is consistent with the data in conventional AF 
transitions.\cite{Killer02} 

To conclude, we report measurements of the local lattice disorder as a
function of annealing time in UCu$_4$Pd.  The XAFS measurements show that the 
main effect of annealing is to decrease the fraction of Pd~atoms on the 
nominally Cu~16$e$ sites, in agreement with the change in lattice 
parameter.\cite{Weber01}
Although the local U-Cu environment is well ordered, even in the unannealed 
samples, we observe a small but statistically significant narrowing of its 
pair-distance distribution on annealing. This leads to the conclusion that
disorder must be included in any complete microscopic theory of NFL properties in
the UCu$_{5-x}$Pd$_x$ system. Still, in order to quantitatively understand
the NFL behavior within a Kondo disorder model, one is forced to assume a 
significant amount of hidden disorder. Alternatively, 
quantum critical fluctuations may play a dominant role for the observed 
NFL behavior. Alternate models that \textit{combine} both the QCP and disorder
ideas,\cite{Castro-Neto98,Grempel99} might be able to fill this gap.
Clearly, more work is required to arrive at a satisfactory theoretical description
of the UCu$_{5-x}$Pd$_x$ system.

We thank W. W. Lukens, D. A. Shaughnessy and R. A. Wilson for assistance
in collecting the XAFS data.  We also thank E. D. Bauer, N. B\"uttgen, 
A. H. Castro Neto, D. E. MacLaughlin, E. Miranda, R. Movshovich, and G. R. Stewart
for many useful discussions.  This work was supported by the Director, Office of 
Science, Office of Basic Energy Sciences (OBES), Chemical Sciences, Geosciences, 
and Biosciences Division, U.S. Department of Energy (DOE)
under Contract No.  AC03-76SF00098,
and by SFB~484 of the Deutsche Forschungsgemeinschaft (DFG). 
Data were collected at the SSRL, which is operated by the DOE/OBES.

\bibliography{/home/hahn/chbooth/papers/bib/bibli}

\end{document}